\documentclass[twocolumn,prd,nofootinbib,showpacs,preprintnumbers,floatfix]{revtex4}
\usepackage[dvips]{graphicx} 
\usepackage{subfigure}
\newcommand{\La}{\mathcal{L}}
\newcommand{\PgP}{\bar\psi\gamma^\mu\psi}
\newcommand{\tensorbilinear}{\bar\psi\frac{i}{2}(\gamma_\mu
\buildrel\rightarrow\over\partial_\nu -
\gamma_\mu\buildrel\leftarrow\over\partial_\nu)\psi}
\begin{document}

\title{Spontaneous Breaking of Lorentz Invariance}
\author{Alejandro Jenkins}
\email{jenkins@theory.caltech.edu}
\affiliation{California Institute of Technology, Pasadena, CA 91125, USA}
\date{Nov. 2003; published in Phys. Rev. D. {\bf 69}:105007, 2004}
\preprint{CALT-68-2470}
\preprint{hep-th/0311127}

\begin{abstract}
We describe how a stable effective theory in which particles of the
same fermion number attract may spontaneously break Lorentz invariance
by giving a non-zero fermion number density to the vacuum (and therefore
dynamically generating a chemical potential term).  This mechanism
yields a finite vacuum expectation value $\langle \bar\psi\gamma^\mu\psi \rangle$,
which we consider in the context of proposed models that require such
a breaking of Lorentz invariance in order to yield composite degrees
of freedom that act approximately like gauge bosons.  We also make
general remarks about how the background source provided by
$\langle \bar\psi\gamma^\mu\psi \rangle$ could relate to work on signals of Lorentz
violation in electrodynamics.
\end{abstract}

\maketitle

\section{Introduction}

Lorentz invariance (LI), the fundamental symmetry of Einstein's
special relativity, states that physical results should not change
after an experiment has been boosted or rotated.  In recent years, and
particularly since the publication of work on the possibility of
spontaneously breaking LI in bosonic string field theory
\cite{strings}, there has been considerable interest in the prospect
of violating LI.  More recent motivations for work on Lorentz
non-invariance have ranged from the explicit breaking of LI in the
non-commutative geometries that some have proposed as descriptions of
physical space-time (see \cite{Madore} and references therein), and in
certain supersymmetric theories considered by the string community
\cite{OoguriVafa}, to the possibility of explaining puzzling cosmic
ray measurements by invoking small departures from LI
\cite{ColemanGlashow} or modifications to special relativity itself
\cite{MagueijoSmolin,Amelino}.  It has also been suggested that
anomalies in certain chiral gauge theories may be traded for
violations of LI and CPT \cite{Klinkhamer}.\footnote{This is far from
a thorough account of the rich scientific literature on Lorentz
non-invariance.  Extensions of the standard model have been proposed
which are meant to capture the low-energy effects of whatever new
high-energy physics (string theory, non-commutative geometry, loop
quantum gravity, etc.) might be introducing violations of LI
\cite{SME}.}

Our own interest in the subject began with a recent proposal
\cite{KrausTomboulis} for addressing the cosmological constant problem
(i.e., how to explain the flatness or near flatness of the Universe
without unnaturally fine tuning the parameters of our quantum
theories) by reviving an old idea for generating composite
force-mediating particles \cite{Bjorken1}.  This sort of mechanism
depends on the spontaneous breaking of LI.  In the following section
of this paper we shall discuss this idea and address some problems
related to obtaining the required LI breaking in the manner that has
been proposed.

This leads us to investigate the question of how a reasonable quantum
field theory {\it might} spontaneously break LI.  Borrowing from some
old theoretical work \cite{NJL, selfconsistent} as well as from the
recent research into color superconductivity \cite{colorsuperconduct1,
colorsuperconduct2, Wetterich}, we argue for the existence of theories
with Lorentz invariant bare Lagrangians where the formation of a
condensate of particles of the same fermion number is energetically
favorable, leading to a non-Lorentz invariant vacuum expectation value
(VEV) $\langle \PgP \rangle \neq 0$.

This breaking of LI can be thought of conceptually as the introduction
of a preferred frame: the rest frame of the fermion number density.
If some kind of gauge coupling were added to the theory without
destroying this LI breaking, the fermion number density would also be
a charge density, and the preferred frame would be the rest frame of a
charged background in which all processes are taking place.  This
allows us to make some very general remarks on the resulting
LI-violating phenomenology for electrodynamics and on experimental
limits to our non-Lorentz invariant VEV.  Most of the work in this
area, however, is left for future investigation.

\section{Emergent Gauge Bosons}

In 1963, Bjorken proposed a mechanism for what he called the
``dynamical generation of quantum electrodynamics'' (QED)
\cite{Bjorken1}.  His idea was to formulate a theory which would
reproduce the phenomenology of standard QED, without invoking local
$U(1)$ gauge invariance as an axiom.  Instead, Bjorken proposed
working with a self-interacting fermion field theory of the form
\begin{equation}
\La = \bar\psi(i\partial\!\!\!/-m)\psi - \lambda(\PgP)^2 .
\label{bjorken}
\end{equation}
Bjorken then argued that in a theory such as that described by
Eq. (\ref{bjorken}), composite ``photons'' could emerge as Goldstone
bosons resulting from the presence of a condensate that spontaneously
broke Lorentz invariance.

Bjorken's idea might not seem attractive today, since a theory such as
Eq. (\ref{bjorken}) is not renormalizable, while the work of 't Hooft
and others has demonstrated that a locally gauge invariant theory can
always be renormalized \cite{tHooft}.  There would not appear to be,
at this stage in our understanding of fundamental physics, any
compelling reason to abandon local gauge invariance as an axiom for
writing down interacting quantum field theories.\footnote{We do know
that in the 1980's Feynman regarded Bjorken's proposal as a
serious alternative to postulating local gauge invariance.  For
enlightening treatments of the principle of gauge invariance and its
historical role in the development of modern physical theories, see
\cite{ORaifStraumann,JacksonOkun}.}

Furthermore, the arguments for the existence of a LI-breaking
condensate in theories such as Eq. (\ref{bjorken}) have never been
solid.  (For Bjorken's most recent revisiting of his proposal, in the
light of the theoretical developments since 1963, see \cite{Bjorken2}).

In 2002 Kraus and Tomboulis resurrected Bjorken's idea for a
different purpose of greater interest to contemporary theoretical
physics: solving the cosmological constant problem
\cite{KrausTomboulis}.  They proposed what Bjorken might call
``dynamical generation of linearized gravity.''  In this scenario a
composite graviton would emerge as a Goldstone boson from the
spontaneous breaking of Lorentz invariance in a theory of
self-interacting fermions.

Being a Goldstone boson, such a graviton would be forbidden from
developing a potential and the existence of exact solutions with
constant matter fields and a massless graviton would be assured.  Then
it would no longer be necessary to fine-tune the cosmological constant
parameter in order to obtain a flat or nearly flat spacetime,
providing a possible solution to a problem that plagues all mainstream
theories of quantum gravity.\footnote{In the bargain, this scheme
would appear to offer an unorthodox avenue to a renormalizable quantum
theory of linearized gravity, because the fermion self-interactions
could be interpreted as coming from the integrating out, at low
energies, of gauge bosons that have acquired large masses via the
Higgs mechanism, so that linearized gravity would be the low energy
behavior of a renormalizable theory.}

In \cite{KrausTomboulis}, the authors consider fermions coupled to
gauge bosons that have acquired masses beyond the energy scale of
interest.  Then an effective low energy theory can be obtained by
integrating out those gauge bosons.  We expect to obtain an effective
Lagrangian of the form:
\begin{eqnarray}
\La &=&
\bar\psi(i\partial\!\!\!/-m)\psi+\sum_{n=1}^\infty\lambda_{n}(\PgP)^{2n}
\nonumber \\ && +\sum_{n=1}^\infty\mu_{n} \left[ \tensorbilinear \right]^{2n} +
\,\ldots
\label{bothbilinears}
\end{eqnarray}
where we have explicitly written out only two of the power
series in fermion bilinears that we would in general expect to get
from integrating out the gauge bosons.

One may then introduce an auxiliary field for each of these fermion
bilinears.  In this example we shall assign the label $A^\mu$ to the
auxiliary field corresponding to $\PgP$, and the label $h^{\mu\nu}$ to
the field corresponding to $\tensorbilinear$.  It is possible to write
a Lagrangian that involves the auxiliary fields but not their
derivatives, so that the corresponding algebraic equations of motion
relating each auxiliary field to its corresponding fermion bilinear
make that Lagrangian classically equivalent to
Eq. (\ref{bothbilinears}).  In this case the new Lagrangian would be
of the form
\begin{eqnarray}
\La ' &=& (\eta^{\mu\nu} + h^{\mu\nu})\tensorbilinear -
\bar\psi(A\!\!\!/ + m)\psi + \ldots \nonumber \\ && -V_A(A^2)
-V_h(h^2) + \ldots
\label{auxiliary}
\end{eqnarray}
where $A^2 \equiv A_\mu A^\mu$ and $h^2 \equiv h_{\mu \nu}
h^{\mu \nu}$.  The ellipses in Eq. (\ref{auxiliary}) correspond to
terms with other auxiliary fields associated with more complicated
fermion bilinears that were also omitted in Eq. (\ref{bothbilinears}).

We may then imagine that instead of having a single fermion species we
have one very heavy fermion $\psi_1$ and one lighter one $\psi_2$.
Since Eq. (\ref{auxiliary}) has terms that couple both fermion species
to the auxiliary fields, integrating out $\psi_1$ will then produce
kinetic terms for $A^\mu$ and $h^{\mu\nu}$.

In the case of $A^\mu$ we can readily see that since it is minimally
coupled to $\psi_1$, the kinetic terms obtained from integrating out
the latter must be gauge invariant (provided a gauge invariant cutoff
is used).  To lowest order in derivatives of $A^\mu$, we must then get
the standard photon Lagrangian $-\frac{1}{4}F_{\mu\nu}^2$  (where
$F_{\mu\nu} \equiv \partial_\mu A_\nu - \partial_\nu A_\mu$).  Since
$A^\mu$ was also minimally coupled to $\psi_2$, we then have, at low
energies, something that has begun to look like QED.

If $A^\mu$ has a non-zero VEV, LI is spontaneously broken, producing
three massless Goldstone bosons, two of which may be interpreted as
photons (see \cite{KrausTomboulis} for a discussion of how the exotic
physics of the other extraneous ``photon'' can be suppressed).  The
integrating out of $\psi_1$ and the assumption that $h^{\mu\nu}$ has a
VEV, by similar arguments, yield a low energy approximation to
linearized gravity.

Fermion bilinears other than those we have written out explicitly in
Eq. (\ref{bothbilinears}) have their own auxiliary fields with their
own potentials.  If those potentials do not themselves produce VEV's
for the auxiliary fields, then there would be no further Goldstone
bosons, and one would expect, on general grounds, that those extra
auxiliary fields would acquire masses of the order of the
energy-momentum cutoff scale for our effective field theory, making
them irrelevant at low energies.

The breaking of LI would be crucial for this kind of mechanism, not
only because we know experimentally that photons and gravitons are
massless or very nearly massless, but also because Weinberg and Witten
have shown that a Lorentz invariant theory with a Lorentz invariant
vacuum and a Lorentz covariant energy-momentum tensor does not admit a
composite graviton \cite{WeinbergWitten}.

Let us concentrate on the simpler case of the auxiliary field $A^\mu$.
For the theory described by Eq. (\ref{auxiliary}), the equation of
motion for $A^\mu$ is
\begin{equation}
\frac{\partial \La'}{\partial A_{\mu}} = -\PgP - V'(A^2) \cdot 2
A^{\mu} = 0.
\label{algebraic}
\end{equation}

Solving for $\PgP$ in Eq. (\ref{algebraic}) and substituting into both
Eq. (\ref{bothbilinears}) and Eq. (\ref{auxiliary}) we see that the
condition for the Lagrangians $\La$ and $\La'$ to be classically
equivalent is a differential equation for $V(A^2)$ in terms of the
coefficients $\lambda_{n}$:
\begin{equation}
V(A^2) = 2 A^2 [V'(A^2)] -  \sum_{n=1}^{\infty}\lambda_{n} 2^{2n}
A^{2n} [V'(A^2)]^{2n}.
\label{differen}
\end{equation}

It is suggested in \cite{KrausTomboulis} that for some values of
$\lambda_n$ the resulting potential $V(A^2)$ might have a minimum away
from $A^2=0$, and that this would give the LI-breaking VEV needed.  It
seems to us, however, that a minimum of $V(A^2)$ away from the origin
is not the correct thing to look for in order to obtain LI breaking.
The Lagrangian in Eq. (\ref{auxiliary}) contains $A^\mu$'s not just in
the potential but also in the ``interaction'' term $A_\mu \PgP$, which
is not in any sense a small perturbation as it might be, say, in QED.
In other words, the classical quantity $V(A^2)$ is not a useful
approximation to the quantum effective potential for the auxiliary
field.

In fact, regardless of the values of the $\lambda_{n}$,
Eq. (\ref{differen}) implies that $V(A^2=0)=0$, and also that at any
point where $V'(A^2)=0$ the potential must be zero.  Therefore, the
existence of a classical extremum at $A^2=C \neq 0$ would imply that
$V(C)=V(0)$, and unless the potential is discontinuous somewhere, this
would require that $V'$ (and therefore also $V$) vanish somewhere
between $0$ and $C$, and so on \textit{ad infinitum}. Thus the
potential $V$ cannot have a classical minimum away from $A^2=0$,
unless the potential has poles or some other discontinuity.

A similar observation applies to any fermion bilinear for which we
might attempt this kind of procedure and therefore the issue arises as
well when dealing with the proposal in \cite{KrausTomboulis} for
generating the graviton.  It is not possible to sidestep this
difficulty by including other auxiliary fields or other fermion
bilinears, or even by imagining that we could start, instead of from
Eq. (\ref{bothbilinears}), from a theory with interactions given by an
arbitrary, possibly non-analytic function of the fermion bilinear
$F(\mbox{bilinear})$.  The problem can be traced to the fact
that the equation of motion of any auxiliary field of this kind will
always be of the form
\begin{equation}
0 = -(\mbox{bilinear}) - V'(\mbox{field}^2)\cdot 2 \,\mbox{field} .
\label{general}
\end{equation}

The point is that the vanishing of the first derivative of the
potential or the vanishing of the auxiliary field itself will always,
classically, imply that the fermion bilinear is zero.  Classically at
least, it would seem that the extrema of the potential would
correspond to the same physical state as the zeroes of the auxiliary
field.

\section{Nambu and Jona-Lasinio Model (review)}

The complications we have discussed that emerge when one tries to
implement LI breaking as proposed in \cite{KrausTomboulis} do not, in
retrospect, seem entirely surprising. A VEV for the auxiliary field
would classically imply a VEV for the corresponding fermion bilinear,
and therefore a trick such as rewriting a theory in a form like
Eq. (\ref{auxiliary}) should not, perhaps, be expected to uncover a
physically significant phenomenon such as the spontaneous breaking of
LI for a theory where it was not otherwise apparent that the fermion
bilinear in question had a VEV.  Let us therefore turn our attention
to considering what would be required so that one might reasonably
expect a fermion field theory to exhibit the kind of condensation that
would give a VEV to a certain fermion bilinear.

If we allowed ourselves to be guided by purely classical intuition, it
would seem likely that a VEV for a bilinear with derivatives (such as
$\tensorbilinear$) might require nonstandard kinetic terms in the
action.\footnote{Recent theoretical work in cosmology has shown
interest in scalar field theories with such nonstandard kinetic terms.
See, for instance, \cite{k-inflation,phantom,ghost}.}  Whether or not this
intuition is correct, we abandon consideration of such bilinears here
as too complicated.

The simplest fermion bilinear is, of course, $\bar\psi\psi$.  Being a
Lorentz scalar, $\langle \bar\psi\psi \rangle \neq 0$ will not break LI.  This kind
of VEV was treated back in 1961 by Nambu and Jona-Lasinio, who
used it to spontaneously break chiral symmetry in one of the early
efforts to develop a theory of the strong nuclear interactions, before
the advent of quantum chromodynamics (QCD) \cite{NJL}.  It might be
useful to review the original work of Nambu and Jona-Lasinio, as it
may shed some light on the study of the possibility of giving VEV's to
other fermion bilinears that are not Lorentz scalars.

In their original paper, Nambu and Jona-Lasinio start from a
self-interacting massless fermion field theory and propose that the
strong interactions be mediated by pions which appear as Goldstone
bosons produced by the spontaneous breaking of the chiral symmetry
associated with the transformation $\psi \mapsto
\exp{(i\alpha\gamma^5)} \psi$.  This symmetry breaking is produced by
a VEV for the fermion bilinear $\bar{\psi}\psi$.  In other words,
Nambu and Jona-Lasinio originally proposed what, by close analogy to
Bjorken's idea, would be the ``dynamical generation of the strong
interactions.''\footnote{Historically, though, Bjorken was motivated
by the earlier work of Nambu and Jona-Lasinio.}

Nambu and Jona-Lasinio start from a non-renormalizable quantum field
theory with a four-fermion interaction that respects chiral symmetry:
\begin{equation}
\La = i \bar{\psi} \partial\!\!\!/ \psi - \frac{g}{2}[(\PgP)^2 -
(\bar{\psi}\gamma^\mu \gamma^5 \psi)^2] .
\label{NJL}
\end{equation}

\begin{figure}[t]
\begin{center}
\includegraphics[bb=172 558 448 639, clip, width=0.46\textwidth]{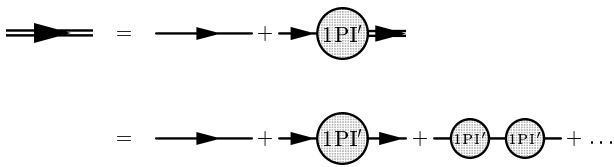}
\caption{\small Diagrammatic Schwinger-Dyson equation.  The double
line represents the primed propagator, which incorporates the
self-energy term.  The single line represents the unprimed propagator.
1PI$'$ stands for the sum of one-particle irreducible graphs with the
primed propagator.}
\label{sde}
\end{center}
\end{figure}

In order to argue for the presence of a chiral symmetry-breaking
condensate in the theory described by Eq. (\ref{NJL}), Nambu and
Jona-Lasinio borrowed the technique of self-consistent-field theory
from solid state physics (see, for instance, \cite{selfconsistent}).
If one writes down a Lagrangian with a free and an interaction part,
$\La = \La_0 + \La_i$, ordinarily one would then proceed to
diagonalize $\La_0$ and treat $\La_i$ as a perturbation.  In
self-consistent field theory one instead rewrites the Lagrangian as
$\La = (\La_0 + \La_s) + (\La_i - \La_s) = \La_0' + \La_i'$, where
$\La_s$ is a self-interaction term, either bilinear or quadratic in
the fields, such that $\La_0'$ yields a linear equation of motion.
Now $\La_0'$ is diagonalized and $\La_i'$ is treated as a perturbation.

In order to determine what the form of $\La_s$ is, one requires that
the perturbation $\La_i'$ not produce any additional self-energy
effects.  The name ``self-consistent-field theory'' reflects the fact
that in this technique $\La_i$ is found by computing a self-energy via
a perturbative expansion in fields that already are subject to that
self-energy, and then requiring that such a perturbative expansion not
yield any additional self-energy effects.

Nambu and Jona-Lasinio proceed to make the ansatz that for
Eq. (\ref{NJL}) the self-interaction term will be of the form $\La_s =
-m \bar{\psi}\psi$.  Then, to first order in the coupling constant
$g$, they proceed to compute the fermion self-energy $\Sigma'(p)$,
using the propagator $S'(p) = i(p\!\!\!/ - m)^{-1}$, which corresponds
to the Lagrangian $\La_0' = \bar{\psi}(i \partial\!\!\!/ - m)\psi$
that incorporates the proposed self-energy term.

\begin{figure}[b]
\begin{center}
\includegraphics[bb=175 620 447 658, clip, width=0.46\textwidth]{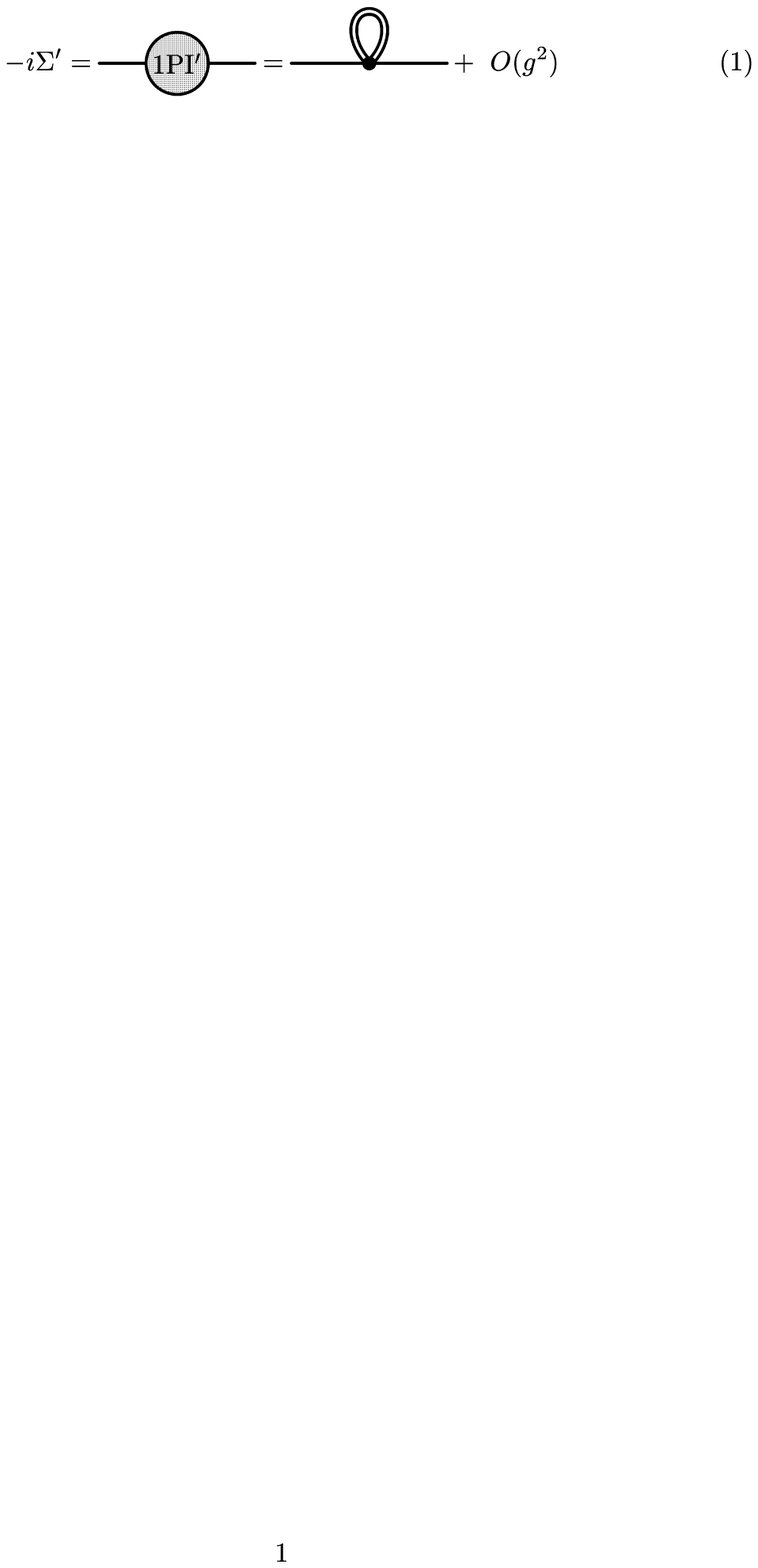}
\caption{\small Diagrammatic equation for the primed self-energy.  We
will work to first order in the fermion self-coupling constant $g$.}
\label{ladder}
\end{center}
\end{figure}

The next step is to apply the self-consistency condition using the
Schwinger-Dyson equation for the propagator:
\begin{equation}
S'(x-y) = S(x-y) + \int d^4 z \, S(x-z) \Sigma'(0) S'(z-y)
\label{Schwinger-Dyson}
\end{equation}
which is represented diagrammatically in Fig. \ref{sde}.
The primes indicate quantities that correspond to a free Lagrangian
$\La_0'$ that incorporates the self-energy term, whereas the unprimed
quantities correspond to the ordinary free Lagrangian $\La_0$.  For
$\Sigma'$ we will use the approximation shown in Fig. \ref{ladder},
valid to first order in the coupling constant $g$.

After Fourier transforming Eq. (\ref{Schwinger-Dyson}) and summing the
left side as a geometric series, we find that the self-consistency
condition may be written, in our approximation, as
\begin{equation}
m = \Sigma'(0) = \frac{g m i}{2\pi^4} \int \frac{d^4 p}{p^2 - m^2 + i
\epsilon} .
\label{NJLselfconsistent}
\end{equation}

If we evaluate the momentum integral by Wick rotation and regularize
its divergence by introducing a Lorentz invariant energy-momentum
cutoff $ p^2 < \Lambda^2 $ we find
\begin{equation}
\frac{2\pi^2 m}{g \Lambda^2} = m \left[ 1 -
\frac{m^2}{\Lambda^2}\log{\left( \frac{\Lambda^2}{m^2} + 1 \right)}
\right] .
\label{NJLselfconsistent2}
\end{equation}

This equation will always have the trivial solution $m=0$, which
corresponds to the vanishing of the proposed self-interaction term
$\La_i$.  But if
\begin{equation}
0 < \frac{2\pi^2}{g\Lambda^2} < 1
\label{NJLcondition}
\end{equation}
then there may also be a non-trivial solution to
Eq. (\ref{NJLselfconsistent2}), i.e., a non-zero $m$ for which the
condition of self-consistency is met.  For a rigorous treatment of the
relation between non-trivial solutions of this self-consistent
equation and local extrema in the Wilsonian effective potential for
the corresponding fermion bilinears, see \cite{Aoki} and the
references therein.

In this model (which from now on we shall refer to as NJL), we see
that if the interaction between fermions and antifermions is
attractive ($g > 0$) and strong enough ($\frac{2\pi^2}{g \Lambda^2} <
1$) it might be energetically favorable to form a fermion-antifermion
condensate.  This is reasonable to expect in this case because the
particles have no bare mass and thus the energy cost of producing them
is small.  The resulting condensate would have zero net charge, as
well as zero total momentum and spin.  Therefore it must pair a
left-handed fermion $\psi_L = \frac{1}{2}(1-\gamma^5)\psi$ with the
antiparticle of a right-handed fermion $\psi_R = \frac{1}{2}
(1+\gamma^5)\psi$, and vice versa.  This is the mass-term
self-interaction $\La_i = -m\bar{\psi}\psi = -m (\bar{\psi}_L \psi_R +
\bar{\psi}_R \psi_L)$ that NJL studies.

After QCD became the accepted theory of the strong interactions, the
ideas behind the NJL mechanism remained useful.  The $u$ and $d$
quarks are not massless (nor is $u$-$d$ flavor isospin an exact
symmetry) but their bare masses are believed to be quite small
compared to their effective masses in baryons and mesons, so that the
formation of $\bar{u}u$ and $\bar{d}d$ condensates represents the
spontaneous breaking of an approximate chiral symmetry.  Interpreting
the pions (which are fairly light) as the pseudo-Goldstone bosons
generated by the spontaneous breaking of the approximate $SU(2)_R
\times SU(2)_L$ chiral isospin symmetry down to just $SU(2)$, proved a
fruitful line of thought from the point of view of the phenomenology
of the strong interaction.\footnote{For a treatment of this subject,
including a historical note on the influence of the NJL model in the
development of QCD, see Chap. 19, Sec. 4 in \cite{Weinberg}.}

Condition Eq. (\ref{NJLcondition}) has a natural interpretation if we
think of the interaction in Eq. (\ref{NJL}) as mediated by massive
gauge bosons with zero momentum and coupling $e$.  For it to be
reasonable to neglect boson momentum in the effective theory, the mass
$\mu$ of the bosons should be $\mu > \Lambda$.  If $e^2 < 2\pi^2$ then
$ g = e^2/\mu^2 < 2\pi^2/\Lambda^2$, which violates
Eq. (\ref{NJLcondition}).  Therefore for chiral symmetry breaking to
happen, the coupling $e$ should be quite large, making the
renormalizable theory nonperturbative.  This is acceptable because the
factor of $1/\mu^2$ allows the perturbative calculations we have
carried out in the effective theory Eq. (\ref{NJL}).  This is why the
NJL mechanism is modernly thought of as a model for a phenomenon
of non-perturbative QCD.

\section{An NJL-Style Argument for Breaking LI}

We have reviewed how NJL formulated a model that exhibited a non-zero
VEV for the fermion bilinear $\bar{\psi}\psi$.  The next simplest
fermion bilinear that we might consider is $\PgP$, which was the one
that Bjorken, Kraus, and Tomboulis considered when they discussed the
``dynamical generation of QED.''  This particular fermion bilinear is
especially interesting because it corresponds to the $U(1)$ conserved
current, and also because it is the simplest bilinear with an odd
number of Lorentz tensor indices, so that a non-zero VEV for it would
break not only LI but also charge (C), charge-parity (CP), and
charge-parity-time (CPT) reversal invariance.  C and CP may not be
symmetries of the Lagrangian, as indeed they are not in the standard
model, but by a celebrated result CPT must be an invariance of any
reasonable theory (see \cite{CPT} and references therein).  This
invariance, however, may well be spontaneously broken, as it would be
by any VEV with an odd number of Lorentz indices.

Before proceeding, however, it may be advisable to try to develop some
physical intuition about what would be required for a fermion bilinear
like $\PgP$ to exhibit a VEV.  If we choose a representation of the
gamma matrix algebra and use it to write out $(\PgP)^2$ for an
arbitrary bispinor $\psi$, we may check that $(\PgP)^2 \geq 0$ for
the choice of mostly negative metric $g^{\mu\nu} =
\mbox{diag}(1,-1,-1,-1)$.  That is, $\PgP$ is timelike.  This has an
intuitive explanation, based on the observation that $\PgP$ is a
conserved fermion-number current density.  Classically a charge
density $\rho$ moving with a velocity $\vec{v}$ will produce a current
$j^\mu = (\rho, \rho \vec{v})$ (in units of $c = 1$).  Therefore the
relativistic requirement that the charge density not move faster than
the speed of light in any frame of reference implies that $j^2 \geq
0$.  Considerations of causality make it natural to expect that
something similar would be true of $\PgP$.

For any time-like Lorentz vector $n^\mu$ it is possible to find a
Lorentz transformation that maps it to a vector $n'^\mu$ with only one
non-vanishing component: $n'^0$.  For a constant current density
$j^\mu$, this means that for $j^\mu$ to be non-zero there must be a
charge density $j^0$, which has a rest frame.  Therefore we only
expect to see a VEV for $\PgP$ if our theory somehow has a vacuum with
a non-zero fermion number density.  The consequent spontaneous breaking
of LI may be seen as the introduction of a preferred reference frame:
the rest frame of the vacuum charge.

In the literature of finite density quantum field theory and of color
superconductivity (see, for instance, \cite{colorsuperconduct1} and
\cite{colorsuperconduct2}), the Lagrangians discussed are explicitly
non-Lorentz invariant because they contain chemical potential terms of
the form $f\cdot\bar{\psi}\gamma^0\psi$ .  This term appears in
theories whose ground state has a non-zero fermion number because, by
the Pauli exclusion principle, new fermions must be added just above
the Fermi surface, i.e., at energies higher than those already
occupied by the pre-existing fermions, while holes (which can be
thought of as antifermions) should be made by removing fermions at
that Fermi surface.  The result is an energy shift that depends on the
number of fermions already present and which has opposite signs for
fermions and antifermions.

The physical picture that emerges is now, hopefully, clearer:  a
theory with a VEV for $\PgP$ is one with a condensate that has non-zero
fermion number.  This means that only theories with some form of
attractive interaction between particles with the same sign in fermion
number may be expected to produce such a VEV.  The situation is
closely analogous to BCS superconductivity \cite{BCS}, in which a
phonon-mediated attractive interaction between electrons allows the
presence of a condensate with non-zero electric charge.  Note that in
the NJL model, the condensate was composed of fermion-antifermion
pairs, and therefore clearly $\langle \bar{\psi}\gamma^0\psi \rangle  = 0$, which
implies $\langle \PgP \rangle = 0$.  It should now be physically clear why a VEV for
$\PgP$ would break not only LI but also C, CP, and CPT.

There is an easy way to write a theory which will have a VEV for a
$U(1)$ conserved current: to couple a massive photon to such a current
via a purely imaginary charge.  To see this, let us write a Proca
Lagrangian for a massive photon field with an external source:
\begin{equation}
\La = -\frac{1}{4}F_{\mu\nu}^2  + \frac{\mu^2}{2}A^2 - j_\mu A^\mu .
\label{Proca}
\end{equation}

The equation of motion for the photon field is
\begin{equation}
\partial_\mu F^{\mu\nu} = j^\nu - \mu^2 A^\nu .
\label{EOMProca}
\end{equation}

At energy scales well below the photon mass $\mu$, the kinetic term
$-F_{\mu\nu}^2 /4$ may be neglected with respect to the mass term
$\mu^2 A^2 /2$.  We may then integrate out the photon at zero momentum
by solving the equation of motion Eq. (\ref{EOMProca}) for the photon
field $A^\mu$ with its conjugate momenta $F^{\mu\nu}$ set to zero, and
substituting the result back into the Lagrangian in Eq. (\ref{Proca}).
The resulting low-energy effective field theory has the Hamiltonian
\begin{equation}
\mathcal H_{\mathrm{effective}} = \frac{j^2}{2 \mu^2} .
\label{Procaeff}
\end{equation}

Nothing interesting happens if the source is a timelike current
density, since in that case Eq. (\ref{Procaeff}) has its minimum at
$j^\mu=0$.  But if we were to make the charge coupling to the photon
imaginary (e.g., $j^\mu = i e \PgP$ for $e$ real), then $j^2$ is
actually always negative (recall that $(\PgP)^2$ is always positive)
and we get a ``potential'' with the wrong sign, so that the energy can
be made arbitrarily low by decreasing $j^2$.  If we make $j^\mu$
dynamical by adding to the Lagrangian terms corresponding to the field
that sets up the current, we might expect, for certain parameters
in the theory, that the energy be minimized for a finite value of
$j^\mu$.

By making the charge purely imaginary, our effective theory at energy
scales much lower than the photon mass $\mu$ will look similar to
Eq. (\ref{NJL}), except that the four-fermion interaction in the
effective Lagrangian will be $e^2(\PgP)^2/2\mu^2$ (with an overall
positive, rather than a negative, sign).  What this means is that
fermions are attracting fermions and antifermions are attracting
antifermions, rather than what we had in NJL (and in QED):
attraction between a fermion and an antifermion.  Condensation, if it
occurs, will here produce a net fermion number, spontaneously breaking
C, CP, and CPT.\footnote{At one point, Dyson argued that such a
theory with attraction between particles of the same fermion number
would be unstable and used this to suggest that perturbative series in
QED might diverge after renormalization of the charge and mass
\cite{Dyson}.  We will address the issue of stability at the end of
this section.}

Let us analyze this situation again more rigorously using
self-consistent field theory methods, following Nambu and
Jona-Lasinio.  For this we consider a fermion field with the usual
free Lagrangian $\La_0 = \bar{\psi}(i\partial\!\!\!/ - m_0)\psi$ and
pose as our self-consistent ansatz:
\begin{equation}
\La_s = -(m - m_0)\bar{\psi}\psi - f \bar{\psi}\gamma^0\psi .
\label{Ansatz}
\end{equation}

The corresponding momentum-space propagator for $\La_0' = \La_0 +
\La_s$ is, therefore,
\begin{equation}
S'(k) = i(k\!\!\!/ - f\gamma^0 - m)^{-1} .
\label{primedpropagator}
\end{equation}

Now let us suppose that the interaction term looks like
\begin{equation}
\La_i = \frac{g}{2}(\PgP)^2 .
\label{ajvinteraction}
\end{equation}

\begin{figure}[t]
\begin{center}
\includegraphics[bb=191 625 434 667, clip,
width=0.46\textwidth]{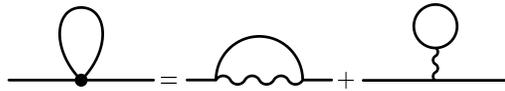}
\caption{\small The four-fermion vertex in the self-interacting theory
may be seen as the sum of two photon-mediated interactions with a
massive photon that carries zero momentum and is coupled to the
fermion via a purely imaginary charge.}
\label{contractions}
\end{center}
\end{figure}

To obtain the Feynman rules corresponding to
Eq. (\ref{ajvinteraction}) we note that this is what we would obtain
in massive QED if we replaced the charge $e$ by $ie$ and the usual
photon propagator by $i g^{\mu\nu}/\mu^2$, with $g = e^2/\mu^2$.  Therefore
to compute the self-energy we will rely on the identity represented in
Fig. \ref{contractions}.  (In QED the second diagram on the right-hand
side of Fig. \ref{contractions} would vanish by Furry's theorem, but
in our case the propagator in the loop will have a chemical potential
term that breaks the C invariance on which Furry's theorem depends.)

To leading order in $g$, the self-energy is
\begin{equation}
\Sigma(0) =  2ig \int \frac{d^4 k}{(2\pi)^4}\frac{3 (k_0 - f) \gamma^0
+ 3 k_i\gamma^i - 2m}{k_0^2-\vec{k}^2-m^2+f^2-2fk_0 + i\epsilon\sigma}
\label{sigmaunsimplified}
\end{equation}
where $\sigma$ (a function of $|\vec{k}|$, $f$, and $m$)
takes values $\pm 1$ so as to enforce the standard Feynman
prescription for shifting the $k^0$ poles:  positive $k^0$ poles are
shifted down from the real line, while negative poles are shifted up.

At first sight it might appear as if the self-energy in
Eq. (\ref{sigmaunsimplified}) could not be used to argue for the
breaking of LI, because the shift in the integration variable $k
\mapsto k' = (k^0 - f,\vec{k})$ would wipe out $f$ dependence.  This,
however, is not the case, as we will see.  We may carry out the $dk^0$
integration, for which we must find the corresponding poles.  These
are located at
\begin{equation}
k_0 = f \pm \sqrt{\vec{k}^2+m^2} .
\label{k0poles}
\end{equation}

From now on, without loss of generality, we will take $f$ to be
positive.  The contour integral which results from closing the $d^0k$
integral of Eq. (\ref{sigmaunsimplified}) in the complex plane will vanish unless
$f < \sqrt{\vec{k}^2+m^2}$, because otherwise both poles in
Eq. (\ref{k0poles}) will lie on the same side of the imaginary axis.
In light of the Feynman prescription used for the shifting of the
poles away from the real axis, it would then be possible to close the
contour at infinity so that there would be no poles in the interior.
The pole shifting prescription, through its effect on the $dk^0$
integral, is what introduces an actual $f$ dependence into the
expression for the self-energy.

By the Cauchy integral formula, we have
\begin{eqnarray}
\Sigma(0) &=& \frac{-g}{4\pi^3} \int d^3 k \left[
\frac{3\sqrt{\vec{k}^2+m^2}\gamma^0  +
2m}{2\sqrt{\vec{k}^2+m^2}}\right. \nonumber \\ &&
\left. \phantom{\frac{1}{1}}
\times \theta (\sqrt{\vec{k}^2+m^2} - f) -\frac{3}{2}\gamma^0 \right]
\label{sigmasimplified}
\end{eqnarray}
where the second term in the right hand side subtracts the
contribution from closing the contour out at infinity in the complex
plane (note the branch cut in the logarithm that results from
computing that part of the contour integral explicitly).  We will
introduce the cutoff $\vec{k}^2 < \Lambda^2$ to make the integral in
Eq. (\ref{sigmasimplified}) finite.\footnote{Carrying out the $dk^0$
integration separately from the spatial integral is legitimate and
useful in light of the form of Eq. (\ref{sigmaunsimplified}), which
does not lend itself naturally to Wick rotation.  But the use of a
non-Lorentz invariant regulator may cause concern that any breaking of
LI we might arrive at could be an artifact of our choice of regulator.
An alternative is to dimensionally regulate
Eq. (\ref{sigmasimplified}) by replacing $d^3 k$ with $d^{d-1} k$.
The resulting equations are more complicated and the dependence on the
range of energies where our non-renormalizable theory is valid is
obscured, but the overall argument does not change.  It is also
possible to multiply the integrand in Eq. (\ref{sigmaunsimplified}) by
a cutoff in Minkowski space $\theta(\Lambda^2 + k^2) = \theta
(\Lambda^2 + k_0^2 - \vec{k}^2)$.  For $\vec{k}^2 < \Lambda^2$ we get
the same result as in Eq. (\ref{sigmasimplified}).  For $\vec{k}^2 >
\Lambda^2$ we must impose the condition that $k_0^2 > \vec{k}^2 -
\Lambda^2$, yielding an additional, rather complicated term which does
not affect the logic of our discussion in this section.  It should be
pointed out that previous work on LI breaking has used 3-momentum
cutoffs in computing self-energies \cite{Soldati}, although in that
case there seems to be a physical interpretation for such a cutoff
which does not apply to the present discussion.  The original work of
Nambu and Jona-Lasinio \cite{NJL} considers cutoffs in Euclidean
4-momentum and in 3-momentum, arriving in both cases at similar
conclusions.}

Note that the Heaviside step function $\theta (\sqrt{\vec{k}^2+m^2} -
f) $ in Eq. (\ref{sigmasimplified}) is always unity if $m > f$, so
that there will be no $f$ dependence at all in
Eq. (\ref{sigmasimplified}) unless $m \leq f$.  Assuming that $m \leq
f$ we have
\begin{eqnarray}
\Sigma(0)&=& \frac{-g}{2\pi^2} \left[-(f^2-m^2)^{3/2}\, \gamma^0
+ m^3\log{(f+\sqrt{f^2-m^2})} \right. \nonumber \\  &&
- m^3 \log{(\Lambda + \sqrt{\Lambda^2+m^2})} \nonumber \\  &&
\left. \vphantom{\frac{0}{1}} + m\Lambda \sqrt{\Lambda^2+m^2} - m
f\sqrt{f^2-m^2} \right] .
\label{sigmaregularized}
\end{eqnarray}

\begin{figure*}[t]
\centering \subfigure[]{\includegraphics[scale=0.4]{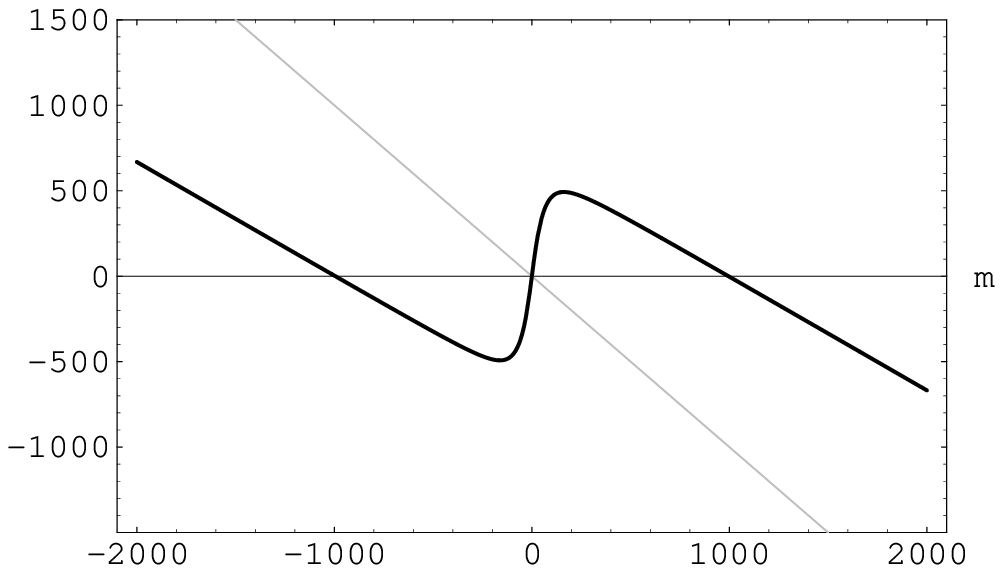}}
\subfigure[]{\includegraphics[scale=0.4]{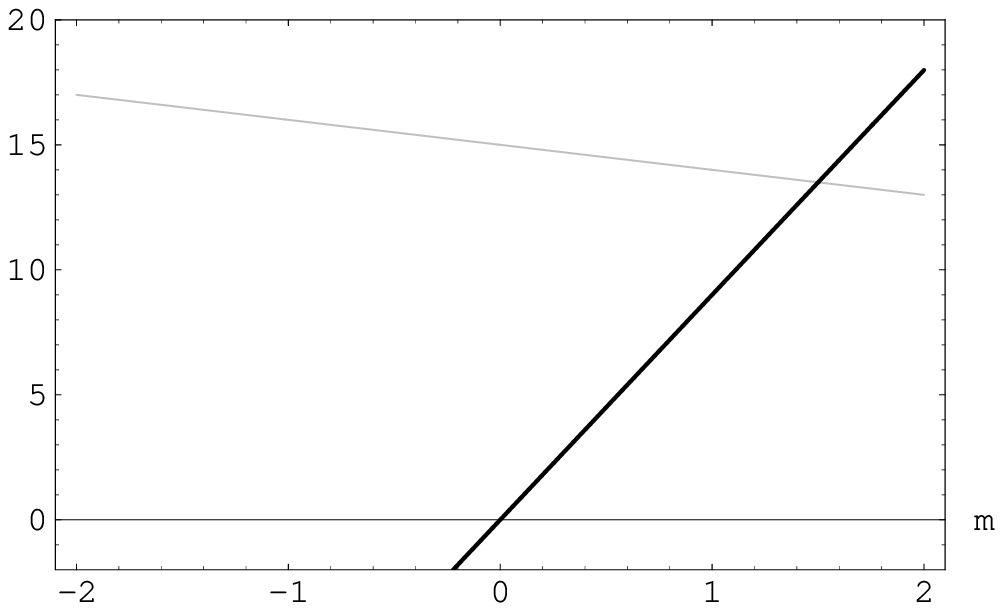}}
\subfigure[]{\includegraphics[scale=0.4]{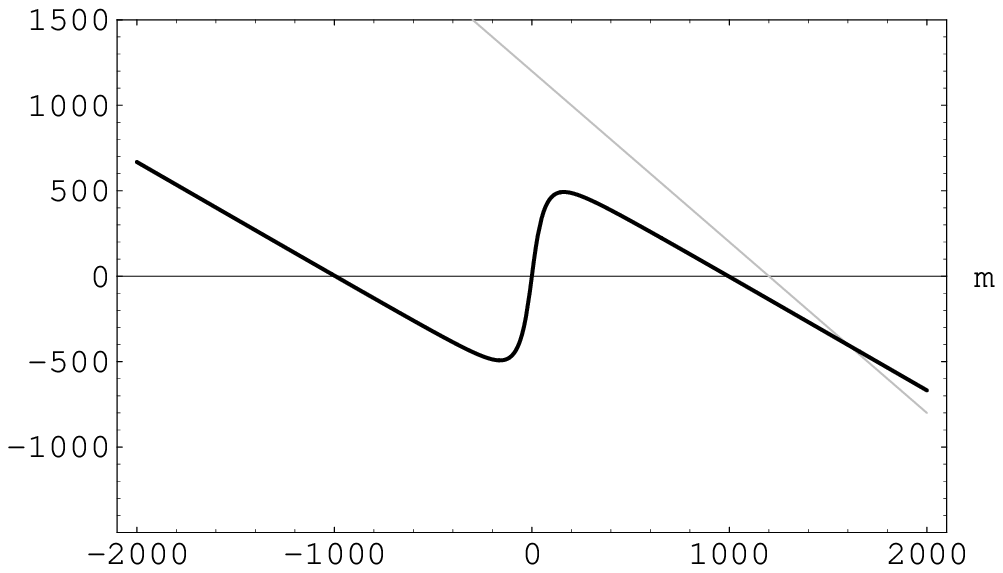}} \\
\subfigure[]{\includegraphics[scale=0.4]{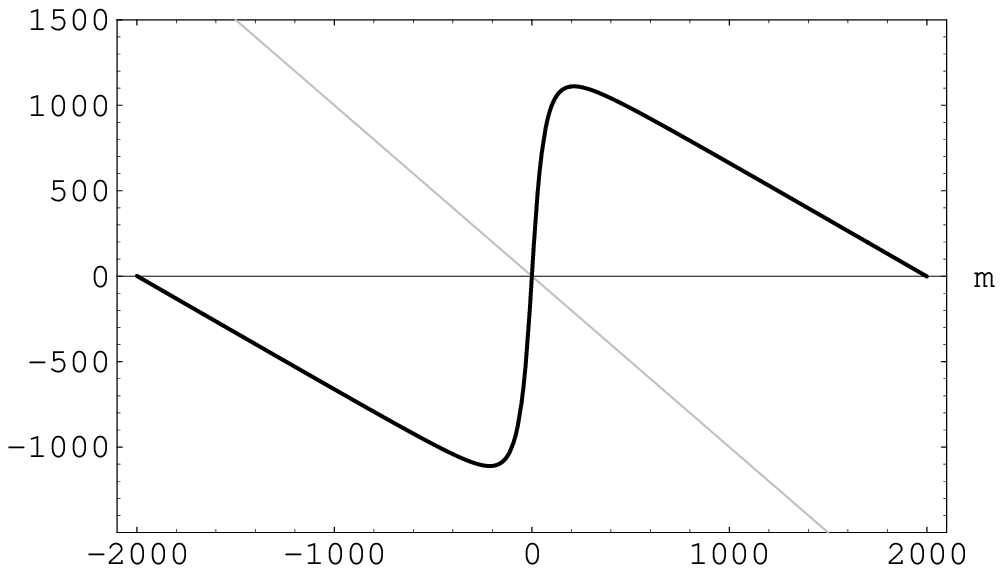}}
\subfigure[]{\includegraphics[scale=0.4]{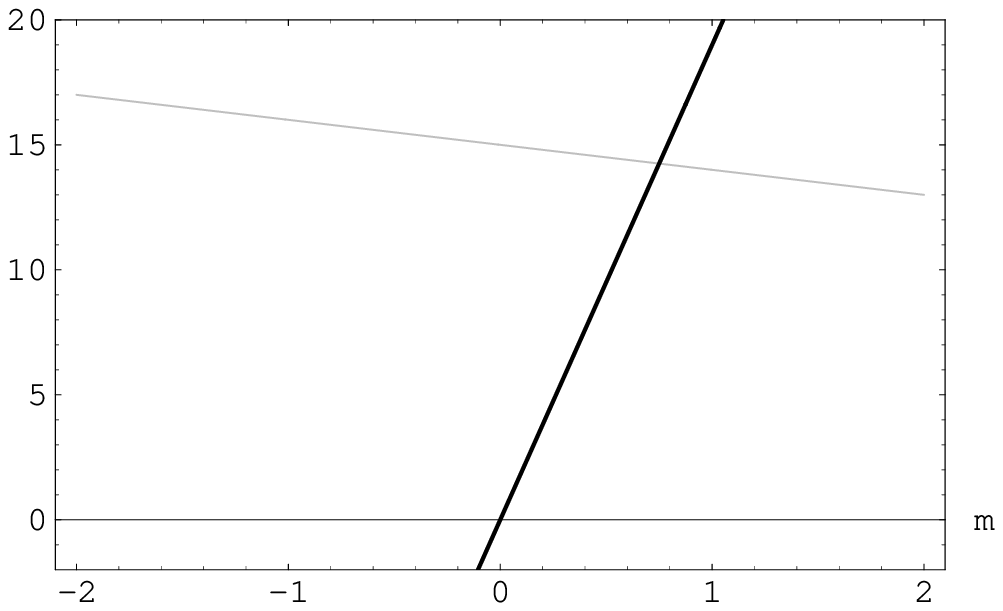}}
\subfigure[]{\includegraphics[scale=0.4]{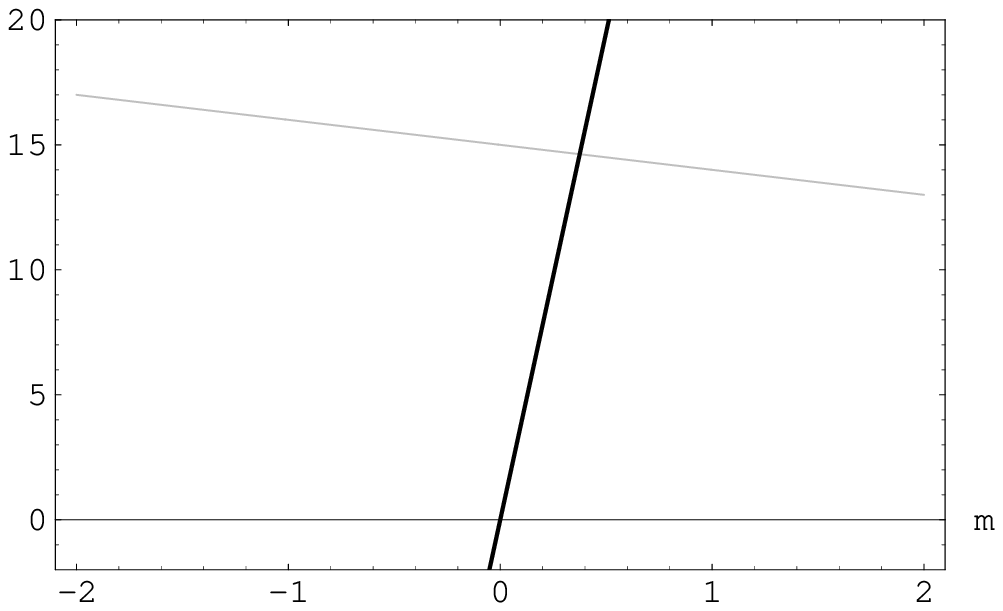}}

\caption{\small Plots of the left-hand side (in gray) and right-hand
side (in black) of equation Eq. (\ref{forgraphs}).  Define $\alpha
\equiv \frac{g}{2\pi^2}$.  For each plot the parameters are: (a)
$\Lambda = 100$, $m_0 = 0$, $\alpha = 0.001$.  (b) $\Lambda = 100$,
$m_0 = 15$, $\alpha = 0.001$.  (c) $\Lambda = 100$, $m_0 = 1200$,
$\alpha = 0.001$.  (d) $\Lambda = 100$, $m_0 = 0$, $\alpha = 0.002$.
(e) $\Lambda = 100$, $m_0 = 15$, $\alpha = 0.002$.  (f)  $\Lambda =
200$, $m_0 = 15$, $\alpha = 0.001$.}
\label{sdesoln1}
\end{figure*}

As before, we use the Schwinger-Dyson equation
Eq. (\ref{Schwinger-Dyson}), and after summing up the right-hand side
as a geometric series, we arrive at the self-consistency condition for
our ansatz Eq. (\ref{Ansatz}):
\begin{eqnarray}
m_0 - m -f\gamma^0 &=& -\Sigma(0) \nonumber \\ &=&
\frac{g}{2\pi^2}\left[-(f^2-m^2)^{3/2} \, \gamma^0 \right. \nonumber \\ && 
+ m^3\log{\left( \frac{f+\sqrt{f^2-m^2}}{\Lambda +
\sqrt{\Lambda^2+m^2}}\right)} \nonumber \\ &&
+ m\Lambda \sqrt{\Lambda^2 + m^2} \nonumber \\ &&
- m f\sqrt{f^2-m^2} \left. \phantom{\frac{0}{1}} \right] .
\label{ajvselfconsistent}
\end{eqnarray}

Clearly Eq. (\ref{ajvselfconsistent}) will not admit a non-trivial
solution $f \neq 0$ unless $g$ is positive, which agrees with our
intuition that the theory must exhibit attraction between particles of
the same fermion number.  The self-consistent condition
Eq. (\ref{ajvselfconsistent}) may be separated into two simultaneous
equations:
\begin{equation}
f  =  \frac{g}{2 \pi^2} (f^2 - m^2)^{3/2}
\label{ajvselfconsistentsep1}
\end{equation}
and
\begin{eqnarray}
m_0 - m & = & \frac{g m}{2 \pi^2} \left[m^2
\log{\left(\frac{f+\sqrt{f^2-m^2}}{\Lambda + \sqrt{\Lambda^2 +
m^2}}\right)} \right. \nonumber \\   && +
\left. \vphantom{\frac{0}{1}} \Lambda \sqrt{\Lambda^2+ m^2} - f
\sqrt{f^2-m^2} \right] .
\label{ajvselfconsistentsep2}
\end{eqnarray}
It is important to bear in mind that
Eqs. (\ref{ajvselfconsistentsep1}) and (\ref{ajvselfconsistentsep2})
were written under the assumption that $f \geq m$.  For $f < m$ the
$f$ dependence of the self-energy in Eq. (\ref{sigmaunsimplified})
disappears.  The trivial, Lorentz invariant solution $f = 0$ to the
self-consistent equations will always be present for any $m$, as
should be the case when spontaneous breaking of a symmetry is observed.

Equation (\ref{ajvselfconsistentsep1}) can be readily solved for $f$
as a function of $m$ (imposing the condition that $f$ be real and
positive), and the resulting $f(m)$ can be substituted into
Eq. (\ref{ajvselfconsistentsep2}) to yield
\begin{eqnarray}
m_0 - m & = & \frac{g m}{2 \pi^2} \left[m^2
\log{\left(\frac{f(m)+\sqrt{f^2(m)-m^2}}{\Lambda + \sqrt{\Lambda^2 +
m^2}}\right)} \right. \nonumber \\  && + \left. \vphantom{\frac{0}{1}}
\Lambda \sqrt{\Lambda^2+ m^2} - f(m) \sqrt{f^2(m)-m^2} \right] .
\label{forgraphs}
\end{eqnarray}

Equation (\ref{forgraphs}) cannot be solved algebraically, but we may
study some of its properties graphically.  In Fig. \ref{sdesoln1} we
have plotted the left-hand side and the right-hand side of
Eq. (\ref{forgraphs}) for various values of the parameters $g$, $m_0$
and $\Lambda$.  As plot (a) illustrates, $m_0 = 0$ implies $m = 0$,
i.e., we cannot dynamically generate both a chemical potential and a
mass term.  For $m=m_0=0$ we have
\begin{equation}
f = \pi\sqrt{2/g} .
\label{fforzerom}
\end{equation}

Plot (b) in Fig. \ref{sdesoln1} shows a $0 < m_0 \ll \Lambda$ for
which the corresponding $m$ will be significantly less than $m_0$.
Plot (c) in the same figure illustrates that a very large $m_0$ is
needed before $m > m_0$, but such solutions are not physically
meaningful because $m_0$ itself is already well beyond the energy
scale for which our effective theory is supposed to hold.  By
comparing plot (b) to plot (e) we may see the effect of increasing $g$
for a given $m_0$ and $\Lambda$.  A comparison of plots (b) and (f)
should illustrate the effect of increasing $\Lambda$ with the other
parameters fixed.

\begin{figure*}[t]
\centering \subfigure[]{\includegraphics[scale=0.38]{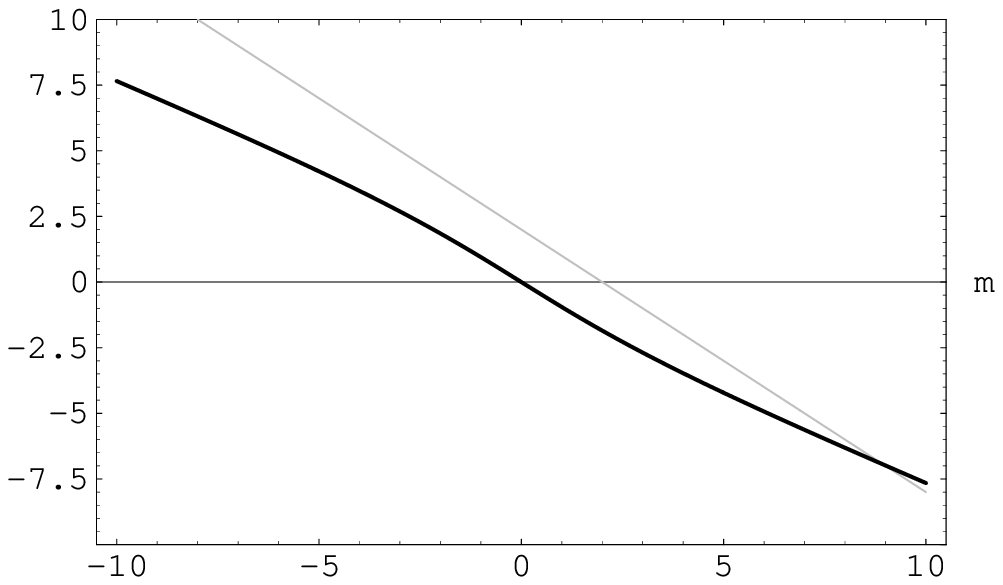}}
\subfigure[]{\includegraphics[scale=0.38]{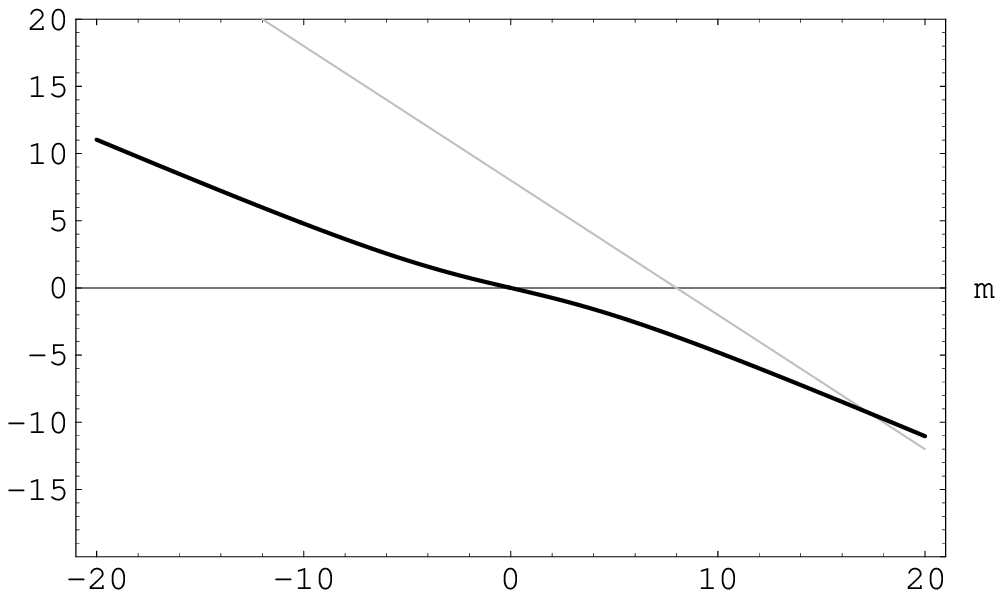}}
\subfigure[]{\includegraphics[scale=0.38]{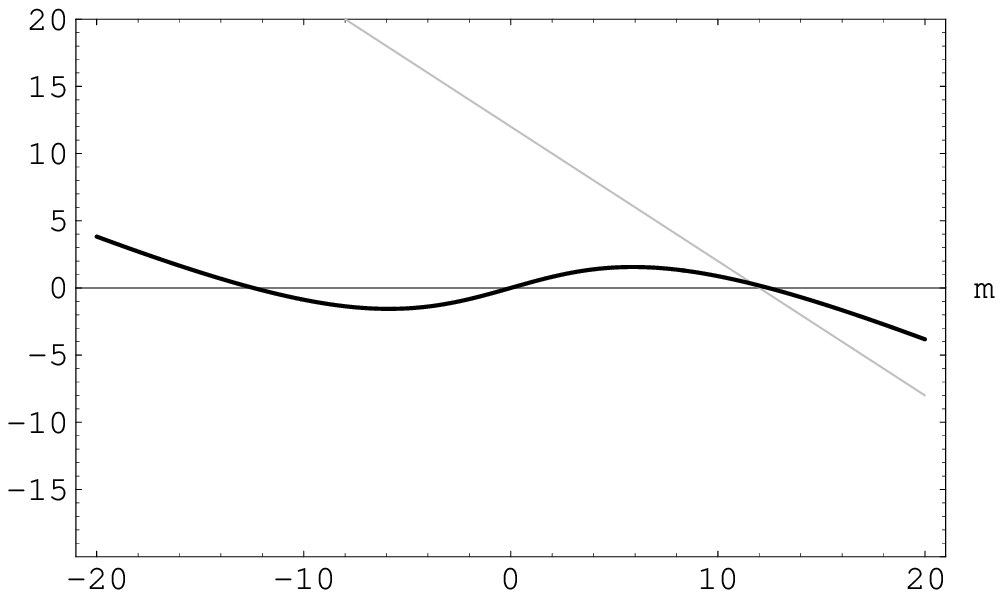}}
\subfigure[]{\includegraphics[scale=0.38]{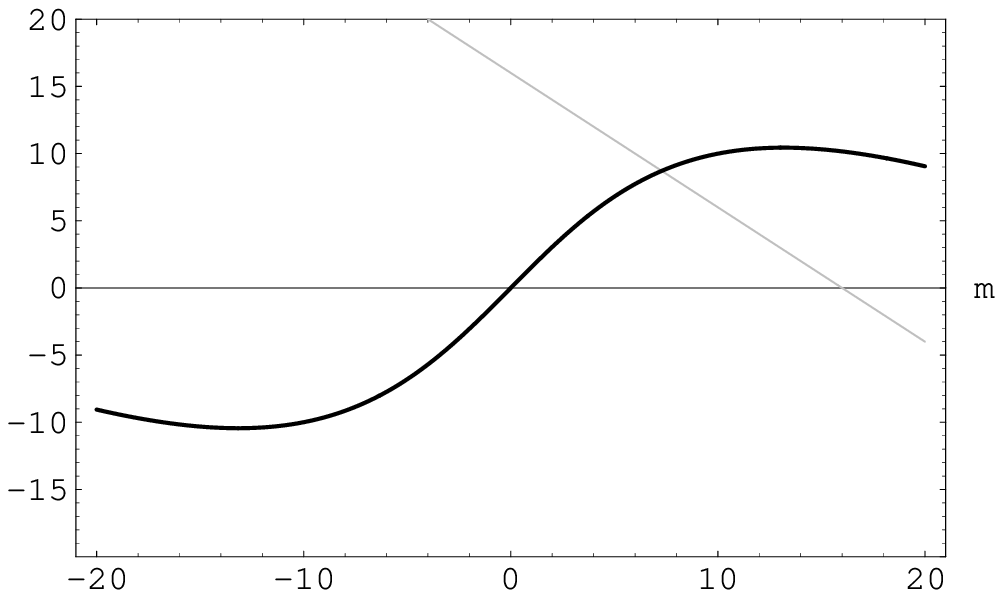}}

\caption{\small Plots of the left-hand side (in gray) and right-hand
side (in black) of equation Eq. (\ref{forgraphs}).  For all of them
$\alpha \equiv \frac{g}{2\pi^2} = 0.01$.  (a) $\Lambda = m_0 = 2$.
(b) $\Lambda = m_0 = 8$.  (c)  $\Lambda = m_0 = 12$.  (d) $\Lambda =
m_0 = 16$.}
\label{sdesoln2}
\end{figure*}

The plots in Fig. \ref{sdesoln2} illustrate the progression, as the
parameter $\Lambda$ is increased for fixed $\alpha$, from an unstable
theory in which bare masses $m_0$ on the order of $\Lambda$ are mapped
to $m > \Lambda$, to a theory that maps such bare masses to $m <
\Lambda$.  Such an analysis of Eq. (\ref{forgraphs}) reveals that the
condition for this mass stability is
\begin{equation}
0 < \frac{2 \pi^2}{g \Lambda^2} < 1
\label{physicalcondition}
\end{equation}
which is reminiscent of the condition
Eq. (\ref{NJLcondition}) for chiral symmetry breaking in the NJL model
(except that now the interaction has the opposite sign).  Combining
Eq. (\ref{physicalcondition}) with Eq. (\ref{fforzerom}) (which was
exact for $m_0$ but may serve approximately for $m_0$ small) we arrive
at the requirement
\begin{equation}
0 < f^2 < \Lambda^2
\label{stability}
\end{equation}
which would surely have to hold if our theory were stable.
Indeed, we may interpret Eq. (\ref{stability}) as saying that if we
pick physically good parameters $g$, $m_0$ and $\Lambda$, we will have
a stable theory with finite chemical potential $f$.  The parameters
for plots (a), (b), (d), (e), and (f) in Fig. \ref{sdesoln1} all give
examples of such stable theories.  As in NJL, the good parameters
involve $g^{-1/2}$ large with respect to $\Lambda$, suggesting that
Eq. (\ref{ajvinteraction}) should be a low-energy approximation to a
non-perturbative interaction of a full renormalizable theory that
allows attraction between particles of the same fermion number sign.

The issue of how the form of the self-consistent equations will depend
on the choice of regulator for the integral in
Eq. (\ref{sigmaunsimplified}) is not an entirely straightforward
matter.  But it seems to be a solid conclusion that, for positive
fermion self-coupling $g$, the solutions to such self-consistent
equations show the presence of LI-breaking vacua.  In the next section
of this paper we offer an alternative approach that strengthens this
conclusion and that sheds further light on the issue of stability.

\section{Consequences for Emergent Photons}

The theory
\begin{equation}
\La = \bar{\psi}(i\partial\!\!\!/ - m_0)\psi + \frac{g}{2}(\PgP)^2
\label{ajv}
\end{equation}
is equivalent to
\begin{equation}
\La' = \bar{\psi}(i\partial\!\!\!/ -A\!\!\!/- m_0)\psi - \frac{A^2}{2g} .
\label{ajv'}
\end{equation}

Since we argued that Eq. (\ref{ajv}) may spontaneously break LI by
giving a finite $\langle \PgP \rangle$, we conclude that $A^\mu$ in Eq. (\ref{ajv'})
would also have a finite VEV, since, by the algebraic equation of
motion,
\begin{equation}
A^\mu = - g \PgP .
\label{bilineartoA}
\end{equation}

This interpretation agrees with the observation that Eq. (\ref{ajv'})
has a vector boson field whose mass term carries the wrong sign if
$g>0$, indicating that the zero-field state is not a good vacuum.  To
find the correct vacuum for the theory we must carry out the path
integral over the fermion field to obtain the effective action
$\Gamma[A]$, and then minimize that quantity.  The field $A^\mu$ is
minimally coupled to $\psi$, so that the computation should proceed as
in QED.  By the Ward identity we do not expect a correction to the
mass term for $A^\mu$, as long as an adequate regulator is used.  But
we do expect to get terms in the effective action that go as $A^4$ and
higher even powers of the auxiliary field.

Since we have reason to believe that QED is stable for any value of
the charge $e$, it therefore seems logical to expect that the
effective action for $A^\mu$ in Eq. (\ref{ajv'}) gives it a finite
time-like VEV, which would imply a finite VEV for $\PgP$ in the theory
of Eq. (\ref{ajv}).  We argued in the previous section that $g$ must
be large for the theory described by Eq. (\ref{ajv}) to be stable.  This too seems natural in
light of Eq. (\ref{ajv'}), because a large $g$ makes the $A^2$ term
small, so that the instability created by it may be easily controlled
by the interaction with the fermions, yielding a VEV for $A^\mu$ that
lies within the energy range of the effective theory.

Armed with Eq. (\ref{ajv'}) it would seem possible to carry out the
program proposed by Bjorken, and by Kraus and Tomboulis, in order to
arrive at an approximation of QED in which the photons are composite
Goldstone bosons.  It is conceivable that a complicated theory of
self-interacting fermions, perhaps one with non-standard kinetic
terms, might similarly yield a VEV for $\tensorbilinear$, allowing the
project of dynamically generating linearized gravity to go forward.
We leave this for future investigation.

\section{Phenomenology of Lorentz Violation by a Background Source}

A separate line of thought that might be pursued from this work
concerns a phenomenology of Lorentz violation in electrodynamics with
a background source.  That is, we might imagine that the fermions of
the universe have some interaction that plays the role of
Eq. (\ref{ajvinteraction}) in giving a VEV to $\PgP$, and that in
addition they have a $U(1)$ gauge coupling (at this stage we have
abandoned the project of producing composite photons).  Then the
$U(1)$ gauge field may interact with a charged background and we would
be breaking LI in electrodynamics by introducing a preferred frame:
the rest frame of the background source.

The possibility of a vacuum that breaks LI and has non-trivial optical
properties has already been investigated in \cite{AndrianovSoldati,
Soldati}. This work, however, deals with significantly more
complicated models, both in terms of the interactions that
spontaneously break LI and of the optical properties of the resulting
vacuum. To obtain a phenomenology for our own simpler proposal, we
consider a free photon Lagrangian of the form
\begin{equation}
\La_0^{\mathrm{photon}} = -\frac{1}{4}F_{\mu\nu}^2 -j_\mu A^\mu
\label{photonlagrangian}
\end{equation}
where $j^\mu = e \langle \PgP \rangle$, thought of as an external source.
The corresponding propagator for the free photon is
\begin{equation}
\langle T\{A^\mu(x)A^\nu(y)\}\rangle = D^{\mu\nu}_F (x-y) + \langle A^\mu(x) \rangle_j
\, \langle A^\nu(y) \rangle_j
\label{photonpropagator}
\end{equation}
where $D^{\mu\nu}(x-y)$ is the connected photon propagator
and $\langle A^\mu(x) \rangle_j$ is the expectation value of $A^\mu$ in the presence
of the external source.

If we take $j^\mu$ constant and naively attempt to calculate the
classical expectation value of $A^\mu$ in the presence of a constant
source by integrating the Green function for electrodynamics, we will
get a volume divergence.  We may attempt to regulate this volume
divergence by introducing a photon mass $\mu$, which gives the result
\begin{equation}
\langle A^\mu(x)\rangle_j = \frac{j^\mu}{\mu^2} .
\label{withphotonmass}
\end{equation}
(It is trivial to check that this is a solution to
$\partial^2 A^\mu + \mu^2 A^\mu = j^\mu$, the wave equation for the
massive photon field with a source.)  This is not satisfactory because
the disconnected term in Eq. (\ref{photonpropagator}) will be
proportional to $\mu^{-4}$ and Feynman diagrams computed with our
modified photon propagator would produce results that depend strongly
on what we took for a regulator.  In fact the mass is physical and
analogous to the effective photon mass first described by the London
brothers in their theory of the electromagnetic behavior of
superconductors \cite{London}.  (Using the language of particle
physics we may say that, in the presence of a $U(1)$ gauge field, the
VEV $\langle \PgP \rangle$ spontaneously breaks the gauge invariance 
and gives a mass to the boson, as in the Higgs mechanism.)

Photons in a superconductor propagate through a constant
electromagnetic source.  In a simplified picture, we may think of it
as a current density set up by the motion of charge carriers of mass
$m$ and charge $e$, moving with a velocity $\vec{u}$.  The proper
charge density is $\rho_0$.  The proper velocity of the charge
carriers is $\eta^\mu = (1, \vec{u})/\sqrt{1-u^2}$.  The source is
then $j^\mu = \rho_0 \eta^\mu = \rho_0 p^\mu / m$, where $p^\mu$ is
the classical energy momentum of the charge carriers.  We may think of
$m$ and $\rho_0$ as deriving from the solutions to the parameters in a
self-consistent equation such as we had in Eq. (\ref{forgraphs}).

The canonical energy momentum $P^\mu$ of the system is $ P^\mu =
m\eta^\mu + eA^\mu = mj^\mu/\rho_0 + eA^\mu$.  As is discussed in the
superconductivity literature (see, for instance, Chap. 8 in
\cite{Kittel}), the superconducting state has zero canonical
energy momentum, which leads to the London equation
\begin{equation}
j^\mu = -\frac{e\rho_0}{m}A^\mu .
\label{Londonright}
\end{equation}
With this $j^\mu$ inserted into the right-hand side of
$\partial^2 A^\mu = j^\mu$ (the wave equation for the photon field in
the Lorenz gauge), we find that we have a solution to the wave
equation of a massive $A^\mu$ with no source and a mass $\mu^2 =
e\rho_0/m$:
\begin{equation}
\partial^2 A^\mu + \frac{e \rho_0}{m} A^\mu = 0 .
\label{Londonwaveequation}
\end{equation}

If we solve for $A^\mu$ in Eq. (\ref{Londonright}) and substitute this
back into Eq. (\ref{photonpropagator}), we get that
\begin{equation}
\langle T\{A^\mu(x)A^\nu(y)\} \rangle = D^{\mu\nu}_F(x-y) + \frac{m^2}{e^2 j^2}j^\mu
j^\nu .
\label{photonpropagatorfinal}
\end{equation}

Notice that if $j^\mu (x)$ is not constant, then Fourier
transformation of the second term in Eq. (\ref{photonpropagatorfinal})
will not yield, in Feynman diagram vertices, the usual energy-momentum
conserving delta function.  Therefore, presumed small violations of
energy or momentum conservation in electromagnetic processes could
conceivably be parametrized by the space-time variation of the
background source.\footnote{This line of thought could connect to work
on LI violation from variable couplings as discussed in
\cite{varyingcoupling}.}

With Eq. (\ref{photonpropagatorfinal}) and a rule for external massive
photon legs, one may then go ahead and calculate the amplitude for
various electromagnetic processes with this modified photon
propagator, and parametrize supposed observed violations of LI (see
\cite{phenomenology1,phenomenology2,phenomenology3}) by $j^\mu$.  If
we can make an estimate of the size of the the mass $m$ of the
background charges, experimental limits on the photon mass ($< 2
\times 10^{-16}$ eV according to \cite{ReviewPP}) will provide a limit
on the VEV of $\PgP$, in light of Eq. (\ref{Londonright}).

\section{Other Possible Consequences of This Mechanism}

There are other consequences of a VEV $\langle \PgP \rangle \neq 0$ on
which we may speculate.  Such a background may have cosmological effects, a line of
thought which might connect, for instance, with \cite{Fried}.  Also,
it is conceivable that such a VEV might have some relation to the
problem of baryogenesis, since it gives the background finite fermion
number and spontaneously breaks CPT, a violation which can ease the
Sakharov condition of thermodynamical non-equilibrium
\cite{baryogenesis}.

It has recently been suggested that the standard model might be
formulated without a Higgs scalar field, by introducing instead
fermion self-interactions which do not destroy the renormalizability
of the theory if there are nonzero UV fixed points under the
renormalization group operation \cite{noHiggs}.  That work, published
after the first manuscript of the present paper had appeared in the
pre-print archive, might well relate to the mechanism we have
described, particularly in light of what was discussed in the previous
section of this paper.

All these tentative ideas are left for possible consideration in the
future.

\section{Conclusions}

We have presented a stable effective theory in which a chemical
potential term is dynamically generated, thus spontaneously breaking
LI (as well as C, CP, and CPT).  The main reasons why this theory
might be interesting are the following: (a) that it might serve as the starting point
for models with emergent gauge bosons, (b) that it could conceivably
point to LI breaking in other more natural theories that share its
fundamental attribute: attraction between particles of the same
fermion number sign (something that is seen in non-Abelian gauge
theories such as QCD, which allows bound states with non-zero baryon
number) and (c) that it produces something that could perhaps interest
those who study the phenomenology of Lorentz violation in
electrodynamics: the breaking of LI by introducing a background source
with its own rest frame.

All of these remain somewhat problematic because (a) our work applies
directly not to the more interesting case of generating emergent
gravitons, but only to photons,  (b) so far we have not been able to
produce models that spontaneously break LI which are significantly
more natural than Eq. (\ref{ajv}), which is a non-renormalizable
theory in which the fermion self-coupling has the opposite sign to
what is obtained by integrating out a heavy $U(1)$ gauge
boson\footnote{Recent work has shown interest in the possibility of
introducing fermion self-couplings that respect non-perturbative
renormalizability \cite{noHiggs}.  This is possible in the presence of
non-zero UV fixed points of the renormalization group.}, and (c) it
remains to be seen whether a phenomenology of electrodynamics with a
background source is of any interest to the effort of explaining the
supposed indications of Lorentz violation in cosmic ray data and other
measurements.  These are all areas that would need to be explored in
order to make more concrete and useful the ideas presented here.

\acknowledgements{The author would like to thank his advisor, M.~B.~Wise, 
for his guidance, and P.~Kraus, J.~D.~Bjorken, J.~Jaeckel, H.~Ooguri, 
K.~Sigurdson, and D.~O'Connell for useful exchanges.  Thanks are also due to
O.~Bertolami and R.~Lehnert for pointing out references to other work
on Lorentz non invariance which were missing from the first draft.
This work was financially supported in part by a R.~A.~Millikan graduate fellowship
from the California Institute of Technology.}

\end{document}